# Tuning the morphology of aerosolised cellulose nanocrystals via controlled aggregation

Daniel Warnes[a*], Jia Hui Lim[b,d], Richard M. Parker[a], Bruno Frka-Petesic[a,c], Ray Freshwater[a], Camila Honorato-Rios[d], Yu Ogawa[b,d], Chiara Giorio[a*], Silvia Vignolini[a,d*]

[a] Yusuf Hamied Department of Chemistry, University of Cambridge, Lensfield Road, Cambridge, CB2 1EW, UK

[b] Université Grenoble Alpes, CNRS, CERMAV, 38000 Grenoble, France

[c] International Institute for Sustainability with Knotted Chiral Meta Matter (WPI-SKCM2), 1-3-1 Kagamiyama, Hiroshima University, Higashi-Hiroshima, Hiroshima, 739-8526, Japan.

[d] Max Planck Institute of Colloids and Interfaces, Am Mühlenberg 1, 14476 Potsdam, Germany

*Correspondence to: dw589@cam.ac.uk; silvia.vignolini@mpikg.mpg.de; chiara.giorio@atm.ch.cam.ac.uk

**Abstract**

Cellulose nanocrystals (CNCs) are polycrystalline, rod-shaped nanoparticles isolated from cellulose, which have attracted increasing attention for a wide variety of applications. While there has been significant research into CNCs in suspensions, hydrogels and films, there have been remarkably few studies that investigated their properties during and after aerosolisation. Here, we studied how aerosolisation impacts the size and morphology of different CNCs suspensions with different surface functionalities. By building a new experimental setup, we observed that colloidally-metastable aqueous CNC suspensions – achieved by carboxylation of the surface hydroxy groups or by exposure to high-intensity ultrasonication – yield large particulates upon aerosolisation under ambient temperatures. In contrast, aqueous suspensions of unfunctionalised CNCs tend to produce, upon aerosolisation, smaller particulates, despite suffering from poor colloidal stability in liquid suspension. Our results demonstrate that both the aerosolisation process itself and the properties of the CNC suspension play a crucial role in determining the final particle size and morphology of CNC-based particles, highlighting the need to consider colloidal stability and surface functionality when designing CNC-based materials for applications involving aerosol delivery or spray-drying.

## 1. Introduction

Cellulose nanocrystals (CNCs) are polycrystalline, high aspect-ratio nanoparticles that can be isolated from natural cellulose via acid hydrolysis and subsequent purification (M.-C. Li et al., 2015). Arising from their desirable intrinsic properties, such as high strength and optical anisotropy (Dufresne, 2013), they are being actively explored for a wide range of applications, including medical biotechnology (Mali & Sherje, 2022), food packaging (Yu et al., 2014), structurally coloured pigments (Droguet et al., 2022; Parker et al., 2022) and scattering enhancers (Guariloff, 2018; Yang et al., 2022; Zhang et al., 2019). The versatility of CNCs is further enhanced by the broad range of surface functionalization strategies available, enabling tailored properties for specific applications. Such modifications include altering their chemistry with functional groups, such as sulfates, azobenzene, block co-polymers or carboxylates (Eskandari et al., 2020; Fraschini et al., 2017; Mendes et al., 2020; Yuan et al., 2018).

The degree of functionalisation and, consequently, the growing number of applications for CNCs has led to an increase in companies producing them, meaning that efficient manufacturing techniques, such as spray drying are essential for generating bulk volumes of dry powder or highly concentrated slurries (Beck et al., 2014; Esparza et al., 2019; Posada et al., 2020). However, numerous studies have shown that the spray-drying of CNCs leads to agglomeration, which results in significant but poorly understood changes in particle morphology. Depending on drying conditions, these can include the formation of microspheres as well as buckled or blistered particles, (Alarcón-Moyano et al., 2023; Di Giorgio et al., 2020; Esparza et al., 2019; Foster et al., 2018). Some studies have investigated the impact of counter-ions on spray dried powder properties. For example, Beck et al. (2012, 2014) describe how the replacement of protons with monovalent cations, such as $Na^+$ ions, as counterions on sulfated CNCs leads to dried powders that are re-dispersible in water upon sonication, with similar properties to the original suspensions. For applications that are strongly sensitive to the size, morphology and surface chemistry of CNCs, such as in the case of their self-assembly in the chiral nematic phase (Frka-Petesic et al., 2023), this is the most commonly adopted method to produce CNC powders. Such small changes can lead to important differences. Other studies have aerosolised CNCs for the purpose of predicting the collection efficiency (Roberts et al., 2019) and fate (Endes et al., 2015) of CNCs in the lungs, finding that an increased aspect ratio leads to an increase in retention by model lung and systems, but presenting no specific nor significant conclusions on how morphology is altered by aerosolisation or how CNC suspension composition drives the aerosol properties (Endes et al., 2015; Roberts et al., 2019).

Here we aim to understand the effects of aerosolisation, at ambient temperature, on a CNC suspension and how this relates to both the initial properties of the nanoparticles and treatments applied to the suspensions. Specifically, we investigate whether an aerosolised droplet containing CNCs undergoes (i) aggregation of individual CNCs into a large particulate or (ii) localised bundling of the CNCs and subsequent breakup into smaller, rod-like particles. In fact, as the local concentration of CNCs within an evaporating droplet increases, the interplay between interparticle interactions and interfacial effects can have strong implications for the resulting CNCs morphologies. To tackle this problem, we study the changes in the morphology and size of the particulates produced after the aerosolisation of a series of CNC suspensions with different surface functionalities. The effect of aerosolisation on unfunctionalised CNCs obtained by hydrochloric acid hydrolysis was studied and compared to those that had undergone an ultrasonication treatment (Parton et al., 2022) or had been surface carboxylated via TEMPO-mediated oxidation (Fraschini et al., 2017). These different types of CNCs are hereafter referred to in this article as Unfun. CNCs (for unfunctionalised CNCs), US-CNCs (for ultrasonicated, unfunctionalised CNCs) and TO-CNCs (for TEMPO-Oxidised CNCs) respectively, whilst the term "CNC" will refer more generally to any cellulose nanocrystal.

Various strategies are used to analyse aerosol particulate morphology. Note that while the term "aerosol" can refer to either liquid droplets or solid particles suspended in a gas (Speight, 2017), for simplicity, here, we will use it, generally to reference the spray generated by the breakup of a CNC suspension into droplets in air, and the subsequent dry particles, after the water has evaporated. The size and morphology of the resultant CNC particulates were determined by a combination of (i) Scanning Mobility Particle Sizing (SMPS), which is an industry-standard and quantifies the electrical mobility diameter distribution of an aerosol sample and (ii) Transmission Electron Microscopy (TEM) on aerosolised CNCs physically sampled and collected, for *post hoc* analysis. Such morphological analysis was then further expanded using scanning nanobeam electron diffraction (SNBED), which allows for the relative orientation of individual crystallites within the aggregated CNC particulates to be determined (Ballu et al., 2024; Lim et al., 2023).

## 2. Methods

*2.1 Production of CNCs by acid hydrolysis*

Unfunctionalised CNCs were obtained from the hydrolysis of cotton (60 g, Whatman No. 1 cellulose filter paper) in hydrochloric acid (860 mL, 3.0 M, Fisher Scientific) under reflux conditions (approximately 110°C) and with strong mechanical agitation. The reaction was quenched after 30 min by dilution with ultrapure Type 1 ice and water (Milli-Q, Millipore), with the simultaneous immersion of the reaction vessel in an ice bath. Soluble cellulose residues and excess acid were removed by three rounds of centrifugation at $20{,}000 \times g$ (30, 20 and 20 min, respectively), with the pellet redispersed in ultrapure water after each round. Excess ions were removed by dialysis against ultrapure water using a MWCO 12–14 kDa membrane. This resulted in a 6.5 wt.% CNC suspension, which corresponds to the "unfunctionalised CNCs" in this study and is the feedstock for any post-treatment.

*2.2 Carboxylation of CNCs*

Carboxylated CNCs were produced by oxidising the above CNC suspension, with 2,2,6,6-tetramethylpiperidinyloxy radical (TEMPO) following a method adapted from the literature (see **Figure S1** (Fraschini et al., 2017; Habibi et al., 2006)). The CNC suspension (10 g, 6.5 wt.%) was diluted with ultrapure water (30 g), to which TEMPO (10 mg) was added under mechanical agitation, until visually dissolved (at least 15 min). NaBr (100 mg) was added to the suspension, followed by NaClO (87%, 13 mL). The ratio of NaClO to cellulose determines the extent of carboxylation and this volume was calculated from calibration against published data (Fraschini et al., 2017). This volume is based on the number of anhydroglucose units and the efficiency of oxidation, with a targeted carboxylation extent of approximately 1200 mmol/kg$_{CNC}$. The pH was monitored by a continuously measuring probe and was initially adjusted to 11.0 through the addition of HCl (1.0 M). When the pH began to drop rapidly (due to consumption of excess base), sodium hydroxide (1.0 M) was added dropwise to maintain the pH in the range 10.0 – 11.0. When the pH stopped decreasing quickly, the reaction was quenched using methanol (30 mL). The sample was purified by centrifugation and dialysis, as described in Section 3.1. The resultant suspension had a concentration of 4.13 wt.%. Fourier transform infrared (FTIR) spectroscopy was used to corroborate that carboxylation had occurred (see **Figure S2**).

*3.3 Conductimetric Titration*

A CNC suspension (3 mg of dry CNC) was diluted with ultrapure water (200 mL final volume). To this was added NaCl (50 µL, 1.0 M) to ensure the suspension is at a sufficient ionic strength and HCl (25 µL, 1.0 M) was added to provide a strong acid contrast to the weak carboxylic acid groups attached to the CNCs. The resulting mixture was titrated against NaOH (1.0 M) with continuous stirring. The concentration of carboxylate groups in the form of moles of carboxylate per gram of CNC was calculated by interpolation of the resulting titration curve (see **Figure S3** and **Equation S1**), according to **Equation 1** (Masruchin & Park, 2015).

$$[COO^-] = \frac{[NaOH](v_{NaOH,2} - v_{NaOH,1})}{m_{CNC}} \qquad (1)$$

Where $[COO^-]$ is the extent of carboxylation in mmol $COO^-$ per gram of CNC dry mass; $[NaOH]$ is the concentration of NaOH in mol $L^{-1}$, $v_{NaOH,1}$ and $v_{NaOH,2}$ refer, respectively, to the first and second equivalence points determined as the total volumes of base needed to neutralise the strong acid and then the weak acid (defined from the intercept of the linear fits of the three distinctive regions (Masruchin & Park, 2015)), and $m_{CNC}$ refers to the dry mass of CNCs.

*2.4 Dynamic Light Scattering (DLS) and Zeta Potential*

A Malvern Panalytical Zetasizer was used to estimate the hydrodynamic size and diffusion coefficient ($D_i$) of never-sprayed CNC suspensions by dynamic light scattering (DLS), as well as their electrophoretic mobility and Zeta potentials. For DLS, 1.2 mL of CNC suspension, diluted to 0.005 wt.% in a 1 mM suspension of NaCl, was input to Malvern cuvettes using a 0.8 µm filtered pipette to prevent dust contamination. For Zeta potential measurements, 35 µL of CNC suspension was added to a zeta cuvette, also diluted at 0.005 wt.% in pure water, without added electrolyte.

*2.5 Aerosol Generation*

In a method adapted from the work of Liu and Lee (1975), a reservoir of CNC suspension (≥400 mL, 0.05 wt.%, mechanical agitation at 80 rpm) was aerosolised with a supply of 3 L/min clean and dry compressed air via a two-fluid, collision type, constant output atomiser (TSI Model 3076). This produced a spray of wet droplets by forcing the liquid suspension through a small orifice, under the pressure of compressed air, and then into a plate where further droplet breakup occurs, as depicted in **Figure 1a**.

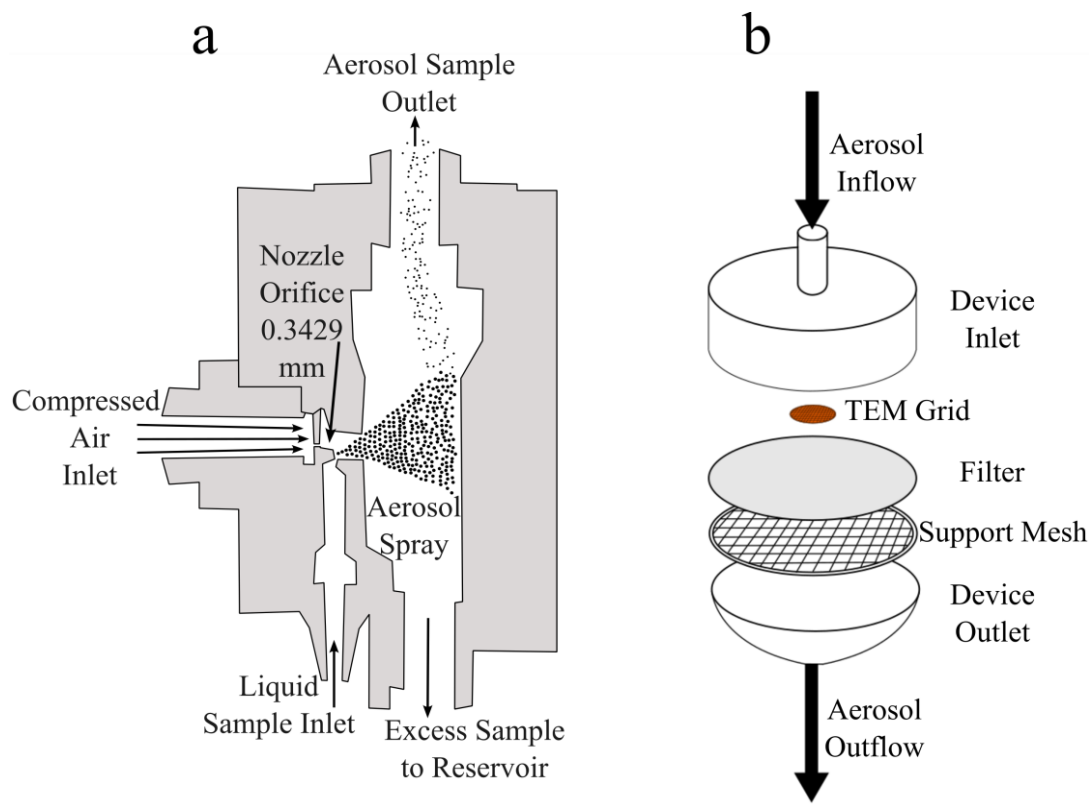

**Figure 1.** Schematic of **(a)** The TSI Constant Output Atomiser 3076 used to aerosolise CNC samples and **(b)** The aerosol particle sampler used to deposit particles onto TEM grids.

**Figure S4** shows the combination of the particle sizing instrumentation in series with the atomiser. The aerosol was dried using a 1 m long diffusion dryer containing silica desiccant beads. To preserve the sample, the aerosol generator was operated in recirculating mode, whereby any excess liquid, not aerosolised by the compressed air, was continuously recirculated to the sample reservoir. This was acceptable because the suspending fluid was ultrapure water and not prone to significant evaporation-induced concentrating effects, which can impact recirculating systems when using volatile solvents (Thermal Systems Inc., 2005). This setup was altered for some experiments to enable ultrasonic treatment of the feedstock suspension during aerosolisation (see **Figure S4**) by using an open reservoir with an ultrasonic probe submerged into the suspension (Blackford & Simons, 1987). This ensured that unfunctionalised CNCs were prevented from aggregating prior to the moment of aerosolisation. A discussion of the aggregation of CNCs in suspension and relevant timescales is also presented after **Figure S4** centring on **Equations S2-S12**.

*2.6 Scanning Mobility Particle Sizing (SMPS)*

The TSI 3080 electrostatic classifier, composed of an $^{85}$Kr bipolar charger and a TSI 3081 Differential Mobility Analyser (DMA), was operated with an inlet sample flow of 0.3 L/min and a sheath air flow of 3.0 L/min. It was used to scan across a range of voltages from 10 to 9581 V, corresponding to an electrical mobility diameter distribution spanning 14 – 673 nm. The mobility diameter is defined as the equivalent diameter of a spherical particle that would have the same electrical mobility as the measured aerosol particle (Lall & Friedlander, 2006). An inertial impactor was used to ensure that large agglomerate particles, exceeding the size range with the potential for equipment damage, would not enter the DMA. The TSI 3775 condensation particle counter (CPC) counts particles fed from the DMA to build a mobility diameter distribution for the aerosol. For every sample, the SMPS system was operated for 80 minutes, corresponding to approximately 36 repeat particle size distributions – this time was selected and tested to ensure any non- steady state phenomena were observable, if present. This also enabled a large enough number of particles to be sampled onto grids for analysis using transmission electron microscopy (TEM).

*2.7 Aerosol Particle Sampling*

SMPS, in isolation, is not ideal for analysing CNCs due to their irregular shape and high aspect ratio (M. Li et al., 2014; Tang et al., 2016). Two aerosol sampling methods were explored, namely filtration (Xiang et al., 2019) and thermophoresis (Patiño et al., 2019). Filtration was ultimately selected due to its simplicity and lower energy requirements. The selected device utilises filtration to deposit sample aerosol particles upon TEM grids (Agar Scientific Cu-C 300 μm grids). As shown in **Figure 1b**, the TEM grids are placed upon a filter membrane (Whatman Cellulose Nitrate, 0.8 μm), which is itself supported by a metal grid within a plastic cassette (SKC Ltd., 25 mm, 2-piece, polypropylene). It is operated vertically such that gravity helps maintain the TEM grid in place. A ball valve and analogue volumetric flowmeter were used to control a bypass line to a sample aerosol flow rate of 0.5 L/min. This was sufficiently low enough that pressure drop and damage to the TEM grid are minimised, whilst sufficiently high that optimal deposition of aerosol particulates upon the grid could be achieved within reasonable time constraints (see **Figure S6**).

*2.8 Transmission Electron Microscopy (TEM)*

A Thermo-Fisher Scientific TALOS TEM was used to visualise the deposited CNC particulates. Liquid samples were ultrasonicated (using a sonication dose of 3000 J ml$^{-1}$ to ensure complete aggregate breakup (Parton et al., 2022)) to enable the visualisation of individual CNCs. Unfunctionalised CNCs aggregate uncontrollably, exacerbated during drying, making it impossible to distinguish between particles. Grids were glow-discharged before deposition for CNC liquid suspensions while both before and after for CNC aerosols. CNC-loaded grids were negatively stained with uranyl acetate (30 μL for 40 s) before being blotted off and allowed to dry. To reduce possible bias in the CNCs selected for imaging, the TEM was scanned across the grid in a cross pattern, taking images at the four corners and the centre (see **Figure S7**). At least 250 particles were analysed, as this was sufficient to produce a representative size distribution (see **Figure S8**).

To extract morphological data from the TEM images of CNCs, a drawing pad was used to manually trace particle contours with sufficient resolution for analysis. The *ImageJ* software was then used with the Shape Filter Plugin (Parton et al., 2022; Wagner & Lipinski, 2013) to extract the relevant morphological parameters: Feret diameter, minimum Feret diameter and aspect ratio (see **Figure S9, Figure S10** and **Equation S13**). Particle size distributions were generated from this information for characterisation and comparison of samples.

*2.9 Scanning Nanobeam Electron Diffraction (SNBED)*

Scanning nanobeam electron diffraction (SNBED) was used to extract crystallographic orientation information from CNCs deposited onto TEM grids within the aerosol sampler. SNBED data was acquired in a low-dose condition optimised for cellulosic crystals as described previously (Lim et al., 2023), using a JEOL 2100F operating at 200 kV and equipped with a NanoMEGAS ASTAR system. The nanobeam configuration consisted of a converged electron probe with a diameter of 15 nm. An electron diffraction (ED) pattern was recorded at every scan position using a MEdipix3 hybrid pixel detector (Cheetah, Amsterdam Scientific Instruments) with a 0.5 ms exposure time. The scan grids were composed of 100 x 100 scan positions with a step size identical to the probe diameter. The diffraction datasets were analysed using the ASTAR® software.

*2.10 Laser Diffraction*

A Sympatec Helium-neon laser for optical spectrometry (HELOS) laser diffraction analyser (19 mm working distance, measurement range 0.1-35 µm) was used to measure the size distribution of the wet droplets in the initial spray. Laser diffraction generates a particle size distribution by measuring the variation in intensity of scattered light as a beam passes through the aerosol, with larger particles scattering light preferentially at smaller angles (forward-scattering). The SMPS experimental set up was modified to enable this measurement by directing the aerosol spray plume generated by the TSI Constant Output Atomiser through the beam of the laser within the HELOS analyser (see **Figure S11**).

3. **Results and Discussion**

The hydrolysis of cellulose to produce CNCs can be achieved using a wide range of acids (Frka-Petesic et al., 2023), most commonly either hydrochloric (such as used in this work) or sulfuric acid. However, whilst the latter simultaneously grafts sulphate half-esters to the surface of the produced CNCs (that can deprotonate into $–O–SO_3^-$ groups), hydrochloric acid does not impart any charged groups (with only –OH present) and thus can be considered "unfunctionalised". As such, to improve the colloidal stability in water, or as a route to introduce additional chemical functionality (e.g. polymer grafting), such "unfunctionalised" CNCs are commonly modified further using TEMPO oxidation (Peyre et al., 2015; Yuan et al., 2018). This reaction results in the selective oxidisation of primary hydroxy groups to carboxylic acids that can deprotonate into $–COO^-$ groups. To better understand the behaviour of CNCs as aerosols, we limited this study to a comparison of HCl-hydrolysed and TEMPO-oxidised CNCs only.

Aqueous suspensions of unfunctionalised and TEMPO-oxidised cellulose nanocrystals (respectively Unfun. CNCs and TO-CNCs) were prepared in water, as described previously in Sections 3.1 and 3.2. Characterisation of these suspensions (**Table 1**) revealed that individual Unfun. CNCs and TO-CNCs exhibit similar physical dimensions (when ultrasonicated prior imaging with transmission electron microscopy as described in section 3.8), but the Unfun. CNCs have a vastly larger hydrodynamic diameter (measured using dynamic light scattering as described in section 3.4). This shows their propensity to aggregate uncontrollably in liquid and highlights the metastability of TO-CNCs in comparison (see **Equations S2-S12**) for a discussion of characteristic aggregation times. Although similar, individual TO-CNCs possess a slightly higher aspect ratio, driven by a lower width, which could arise from delamination and small particle generation being induced during TEMPO-mediated oxidation. Previous studies have confirmed that further chemical and mechanical treatment after initial acid hydrolysis can cause increased fragmentation and chain scission (Benhamou et al., 2014; Levanič et al., 2020; Masruchin et al., 2020; Qin et al., 2011). Furthermore, increased electrostatic forces of repulsion enable lower extents of aggregation (Isogai et al., 2011). The repulsion hypothesis is particularly convincing when considering the greater extent of carboxylation, larger magnitude of zeta potential and smaller hydrodynamic diameter exhibited by TO-CNCs. Confirmation of the presence of carboxylate groups can be found in **Figure S2,** and the extent of carboxylation is examined in **Figure S3**.

**Table 1.** Morphology and surface chemistry of never-sprayed suspensions of unfunctionalised cellulose nanocrystals (Unfun. CNCs) and TEMPO-oxidised cellulose nanocrystals (TO-CNCs), where the ± values represent the standard error of the mean.

| Sample | Dimensions from TEM Analysis | | | Hydrodynamic Diameter (nm) | Zeta Potential (mV) | Carboxylation Extent (mmol/kg) |
| --- | --- | --- | --- | --- | --- | --- |
| | Length (nm) | Width (nm) | Aspect Ratio | | | |
| Unfun. CNC | 151 ± 3 | 14.5 ± 0.2 | 11.3 ± 0.3 | 714 ± 5 | -18.3 ± 0.9 | 83.4 ± 1.0 |
| TO-CNC | 145 ± 3 | 11.4 ± 0.2 | 14.3 ± 0.4 | 160 ± 1 | -42.2 ± 2.8 | 1121 ± 22 |

The Unfunctionalised CNC (Unfun. CNC) suspension and a series of TO-CNC suspensions with differing pH and correspondingly different initial morphologies were aerosolised using the apparatus outlined in **Figure 1** and **Figure S4.** Upon aerosolisation, the electrical mobility diameter of the produced particulates was measured using SMPS while morphological analysis was achieved by sampling the CNC aerosols via deposition onto TEM grids. From qualitative inspection, **Figure 2** shows that after aerosolisation, disk-like CNCs aggregates dominate the grid for TO-CNCs. Moreover, these ellipsoidal objects are significantly larger than for either the initial TO-CNC suspension or for Unfun. CNCs. Since aerosolised CNC-in-water droplets are being formed, we hypothesised that surface tension plays an important role. Surface tension in colloidal systems is influenced by the colloidal stability of particles in that system, which in turn is dependent upon the mutual electrostatic interactions between particles (Anyfantakis et al., 2015; Anyfantakis & Baigl, 2015). To investigate the effect of the electrostatic interactions, hydrochloric acid ([HCl] =1.0 M) and sodium hydroxide ([NaOH] = 1.0 M) were separately added to two samples of TO-CNC suspension (Antoniw et al., 2023). These samples are hereafter referred to as TO-CNC (pH 3) and TO-CNC (pH 11) respectively.

Separately, to access CNCs with different morphology without chemical modification, a sample of Unfun. CNCs that had been pretreated with a high sonication dose (6000 J/ml) were subsequently sonicated continually during aerosolisation. This sample was compared to a never-sonicated aerosol sample from the same feedstock suspension of Unfun. CNCs. Columns 1 and 5 in **Figure 2** show that the sonicated suspension forms significantly larger aerosol particulates (Columns 1 and 5 in **Figure 2**).

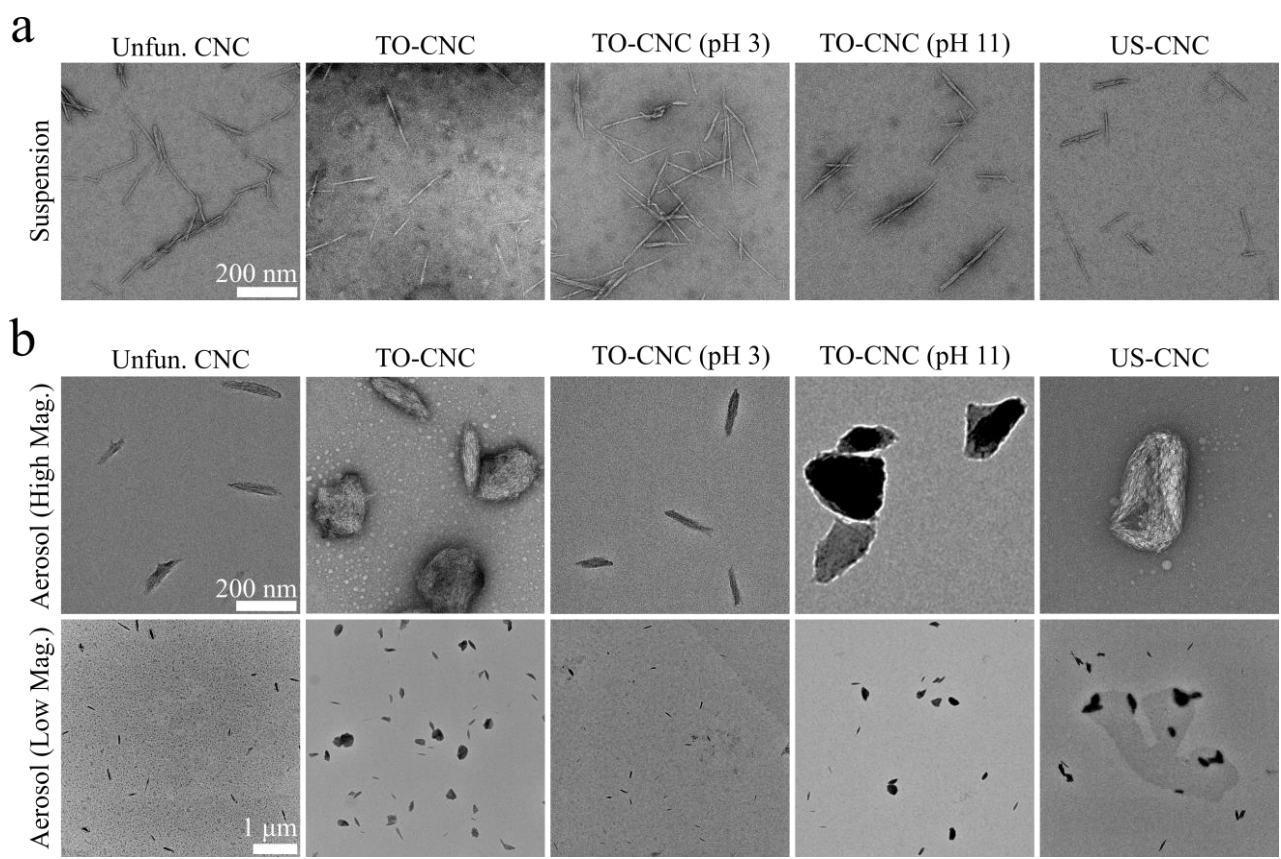

**Figure 2.** TEM images of CNCs **(a)** before and **(b)** after aerosolisation. The aerosolised samples are shown at both high and low magnification (different fields on sample grid) to better visualise the presence of aggregates on the TEM grid.

This TEM analysis allowed for the aerosol samples **(Figure 2b)** to be divided into two distinct categories based on their morphology. The first includes Unfun. CNCs as well as TO-CNCs at pH 3, for which the aerosolised aggregates maintain a high aspect ratio and moderate width. The second group includes US-CNCs, TO-CNCs and TO-CNCs at pH 11, which all exhibit a population of significantly larger aggregates with low aspect ratios and smoother contours. However, it is important to note that even this second group presents significant polydispersity in morphology. As well as the large, rounded aggregates, there still exists a sub-population of smaller particles that exhibit the high aspect ratio morphology, as visualised in the lower-magnification TEM images. This is discussed in greater depth in **Section 4.2**.

To quantify these morphological changes, the outlines of the particles observed in the TEM images were traced (see **Section 3.8** and **Figures S9-S10**). To account for non-uniformities in CNC dimensions, the maximum and minimum Feret diameters of the particles were used to extract the particle length and width, respectively. **Figure 3** presents the particle size distributions for the initial CNC suspensions and for aerosolised CNCs. The top four boxplots confirm that these aerosolised CNC suspensions show a significant increase in length and width when compared to their never-sprayed analogues. This indicates that aggregation is occurring in all samples, albeit to a much greater extent in the initially well-dispersed, TO-CNC sample. Furthermore, the most significant increase is observed in the aerosol particle width, which drives the decrease in aspect ratio.

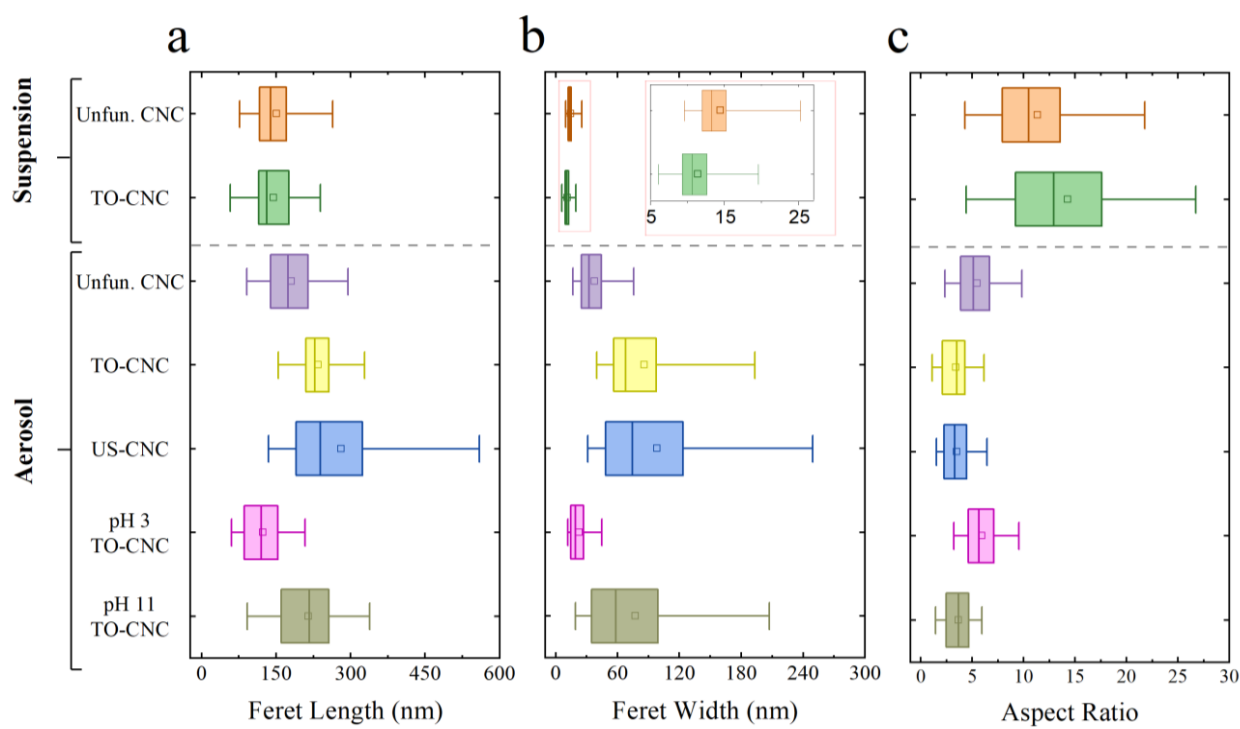

**Figure 3.** Box plots visualising the size and shape distributions in terms of **(a)** Feret length, **(b)** Feret width and **(c)** 2D aspect ratio for CNCs before and after aerosolisation.

We hypothesise that the differences in the aerosol particulate morphology are determined by the colloidal stability of the CNCs, which is much greater for the carboxylated TO-CNCs (due to the larger electrostatic repulsion between these negatively charged nanoparticles, see **Table 1**) than the Unfun. CNCs. If this is the case, then protonation of the carboxylate groups to suppress the surface charge should result in the TO-CNCs behaving in a similar manner to the Unfun. CNCs. To test this effect, the pH was reduced to 3 using 1 mM HCl, which resulted in the aspect ratio of the aerosolised particulates increasing significantly from 3.4 to 5.9, i.e., into the same range as that of the aerosolised Unfun. CNCs (AR = 5.5). Conversely, raising the pH to 11 using 1 mM NaOH results in only cation exchange occurring (Vanderfleet et al., 2022), leading to fully deprotonated TO-CNCs, with a similar ionic strength medium as protonated TO-CNC. In this case, the aspect ratio changed far less for the resultant aerosol particulates (from 3.4 to 3.7). This pH-mediated control over aerosol particulate morphology validates the counter-intuitive prediction that better colloidal stability in suspension results in larger aggregation within the aerosolised droplets.

We observe a similar phenomenon in the US-CNCs, attributed to the increased suspension stability expected for increasingly ultrasonicated CNC suspensions. As the sonication dose increases, the suspension nanocrystal size and hydrodynamic DLS size decrease (Parton et al., 2022). Therefore, the suspensions of US-CNCs can be considered colloidally metastable at least on a certain time scale, despite there being no significant change in the zeta potential (Parton et al., 2022). Thus, aerosolised droplets containing US-CNCs also behave similarly to the colloidally metastable TO-CNCs, forming large, low-aspect ratio aggregates after aerosolisation. However, it can be seen from **Figure 3** that the distribution of aerosolised US-CNCs is considerably more polydisperse.

To understand the mechanism underlying these morphological trends, it is important to understand the properties of the initial "wet" droplets generated by the atomiser. Using laser diffraction (see

**Figure S11**, we observed that droplets generated by this system have a median droplet diameter of 3 µm (ca. 0.014 pL) with distributions that were not significantly different between the Unfun. CNC and TO-CNC aerosols (see **Figure S17a**). This suggests that any difference in the dry particulate morphology is driven by mechanisms occurring after aerosolisation. The distributions are bimodal, with one population occurring with lower frequency in the sub-micron range and a larger population at approximately 4 µm. It is hypothesised that the larger second population arises from collision-induced coalescence of droplets from the first population (Hoffmann & Feingold, 2023).

SMPS was also used to generate particle size distributions for the different aerosols, which were measured after passing through a diffusion drier. The SMPS data (**Figure S17b**) shows that the particulates with the largest mobility diameter are the TO-CNCs and the TO-CNCs at pH 11, with respective mean mobility diameters of 134 and 119 nm. The smallest mobility diameter is observed for the Unfun. CNCs and the TO-CNCs at pH 3, with respective mean mobility diameters of 55 and 53 nm. US-CNCs exhibited a broad, bimodal distribution with peaks at 52 and 143 nm. This ordering is consistent with the trend seen in the TEM data, reported in **Figure 3**. Using the wet droplet size data and the dry particle size data, it was estimated that approximately only 60 individual TO-CNCs would be present within the average aerosol droplet, and perhaps fewer for Unfun. CNCs (see **Equation S16** and the accompanying discussion for a more detailed calculation).

The underlying phenomena for the different aggregation pathways that occur within aerosolised CNC suspensions are proposed in **Figure 4**. First, when an aerosolised droplet containing TO-CNCs is entrained within the airflow of the system, it will begin to dry (**Figure 4a**). Heat and mass transfer occurs, with heat diffusing from the surface of the droplets towards the core and water mass moving down a concentration gradient towards the surface. However, TO-CNCs are colloidally stable in non-acidic conditions due to the strong electrostatic repulsion imparted by the surface carboxylate groups. As such, TO-CNCs will resist this mass transfer due to diffusivity outweighing the rate of shrinkage arising from evaporation ($D_i$ = 5.3E-12 m$^2$ s$^{-1}$ as measured by DLS). This is consistent with the results of an investigation into drying kinetics showing the Péclet number for these aerosolised TO-CNCs to be less than unity ($Pe$ = 0.6, see **Table S1** and the accompanying discussion), which is often used as a critical point for diffusion to dominate advection in drying droplets (Archer et al., 2020; Sobac et al., 2019). As the droplet shrinks, surface tension is maintained (see **Figure S19**) – again due to the colloidal stability of the suspension within the droplet, meaning the TO-CNCs are able to rearrange sufficiently within the drying droplet, such that the droplet-air interface remains unbroken. As drying progresses, the mass fraction of water in the droplet reduces until there is too little water for the TO-CNCs to be suspended, or the system becomes kinetically arrested (see **Figure S22**). At this point, the nanocrystals can no longer repel or diffuse and become cohesive, forming an aggregate. This phenomenon is consistent with literature observations of the spray-dried aggregation of similarly colloidally-stable sulfated CNCs (Esparza et al., 2019; Peng et al., 2012).

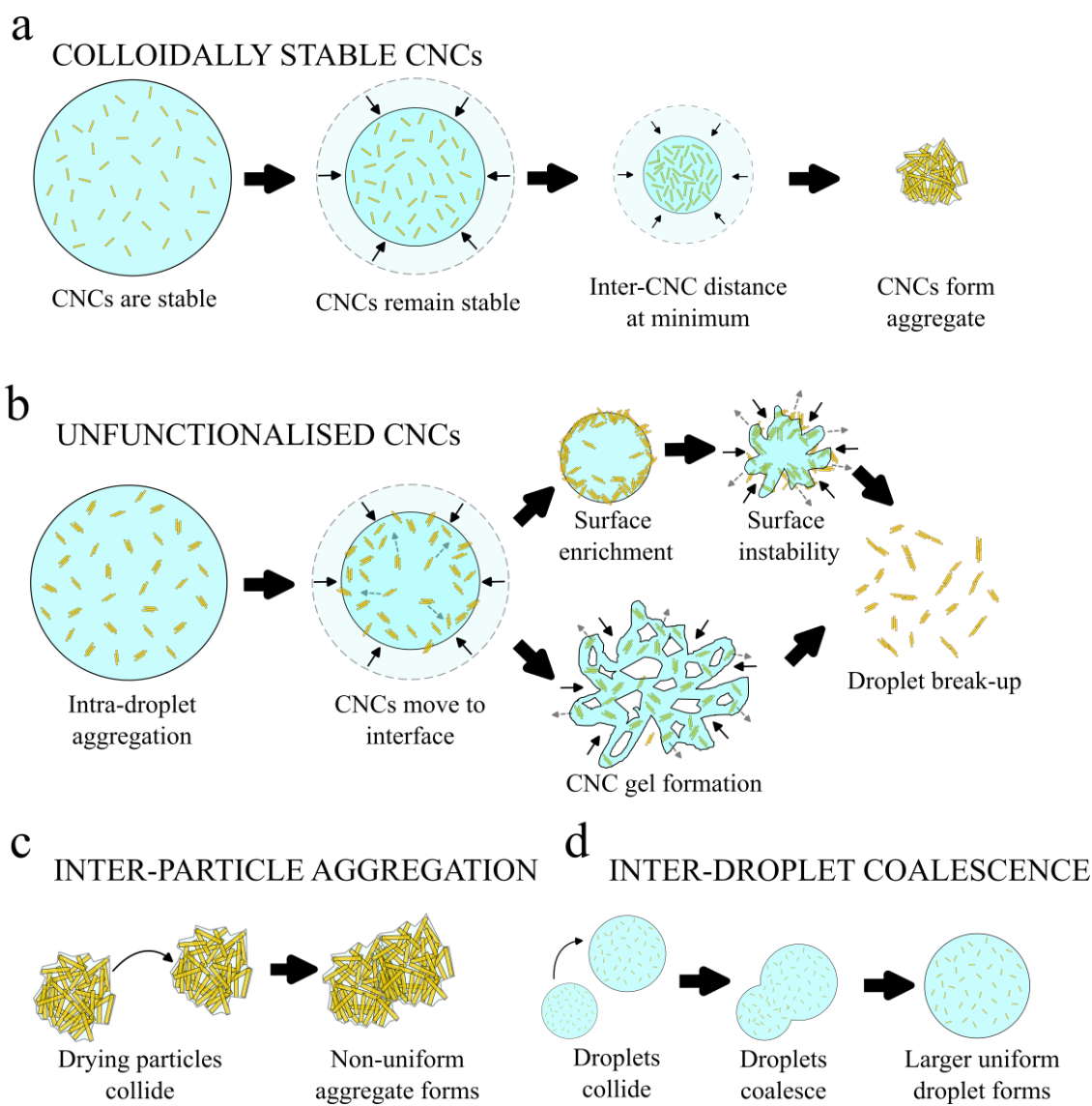

**Figure 4.** Proposed mechanisms for aerosol-induced aggregation for **(a)** colloidally stable CNC suspensions, **(b)** unfunctionalised CNC suspensions, **(c)** non-sphericity caused by collisions between adjacent drying aggregates, and **(d)** formation of larger liquid droplets via coalescence.

A second phenomenon (**Figure 4b**) occurs when Unfun. CNCs are aerosolised. Due to a much lower number of charged surface groups, these Unfun. CNC suspensions are not colloidally metastable and will thus aggregate over time, even in the bulk liquid suspension (Ding et al., 2019). For particularly large, high aspect ratio aggregates (in analogy to spray drying nano-fibrillated cellulose (Peng et al., 2012)) the particles may not even become fully encapsulated within a water droplet during aerosolisation, resulting in a conservation of Unfun. CNC particle morphology from liquid suspension to aerosol. For the aerosolised droplets containing more moderately-sized aggregates, as the droplet begins to dry, heat and mass transfer again begin to occur. As these aggregates diffuse more slowly ($D_i$ = 7.0E-13 m$^2$ s$^{-1}$), relative to the rate of radial droplet shrinkage, the surface becomes enriched in Unfun. CNCs. This is again consistent with an investigation into the Péclet number (***Pe*** = 4.2, see **Table S1**). Upon further drying, the surface of the shrinking droplet becomes increasingly rough, as it is proposed that the high aspect ratio aggregates at the droplet interface might protrude from the drying droplet (cf. (Peng et al., 2012)). At this point, either a decrease in surface tension (due to increased surface roughness as described in **Figure S19-S21**) or the onset of a gel or glassy phase (see **Figure S23**), are hypothesised to cause spontaneous droplet breakup. A combination of both mechanisms is also possible. In the gelation mechanism, capillarity forces at the curvature of surface protrusions overcome the cohesive forces of the gel (Sadek et al., 2015). In the surface enrichment mechanism, a gel may not form, but an outer shell forms as the outer droplet dries faster than the core. Such a shell can cause fragmentation from internal or external pressure (Pramudita et al., 2021), as well as break up due to surface roughness and increased concentration, decreasing the surface tension. Way et al. (2012) suggest that CNCs lacking repulsive interactions with each other are capable of forming gels via a "percolating network of interacting CNC fibres". They suggest that this gel formation can also occur with TO-CNCs, only when TO-CNC suspension pH is reduced, such that carboxylate groups are protonated (forming carboxyl groups), as we ensured in this study. Thus, upon complete loss of water, the broken droplets ultimately form discrete, smaller particles that maintain a high aspect ratio.

Nanocelluloses are being considered as stabilisers in Pickering emulsions (Cherhal et al., 2016; Gestranius et al., 2017; Kalashnikova et al., 2012). The consensus is that surface charge can strongly influence emulsion stability, with Gestranius et al. (2017) stating that the high charge density of carboxylated cellulose nanofibers improves emulsion stability. Kalashnikova et al. (2012) suggest that there is an upper limit to which increasing charge density will positively affect emulsion stability. Cherhal et al. (2016) compare highly-charged sulfated CNCs to more charge-neutral desulfated CNCs and find that desulfated CNCs form fractal aggregates at an oil-water interface that increase roughness and result in lower interfacial coverage with higher porosity. Meanwhile, the sulfated CNCs form in a monolayer, with the more hydrophobic (2 0 0) crystalline plane exposed at the interface (where the carbon-hydrogen groups of CNCs are present). It is possible that the TO-CNCs within an aerosolised droplet in our system are acting on the water-air interface in a similar manner to the sulfated CNCs in the study of Cherhal et al. (2016) on an oil-water interface, forming a smoother, more hydrophobic interface, whilst our Unfun. CNCs act analogously to the desulfated CNCs in that study (see **Figure S21** for scanning electron microscopy images). This is confirmed upon observation of a suspension of colloidally stable Sulf. CNCs, aerosolised using the same setup as described in section 2.5, that form aggregates with rounded, disk-like morphologies, analogous to TO-CNCs (see **Figure S24**).

As an ancillary mechanism, both inter-particle aggregation and inter-droplet coalescence may contribute to the final morphology of the particulates (see respectively **Figures 4c and 4d)**. In the former case, adjacent particulates that are not yet fully dry might collide, leading to larger aggregates with more irregular shapes. Such objects are observed in high abundance in TEM analysis, particularly in TO-CNCs, TO-CNCs at pH 11 and US-CNCs. Conversely, if collisions occur when the volume fraction of water is still high, the droplets can coalesce and will likely dry to form a larger, uniformly shaped aggregate, if suspension is colloidally stable.

Scanning nanobeam electron diffraction (SNBED) was used to understand the morphology and organisation of the aerosol particles at the sub-particulate scale. The aforementioned TEM analysis showed that several discrete morphologies of aerosol particulates were produced upon aerosolisation of cellulose nanocrystal suspensions. These can be subdivided based on their geometry as well as their constituent fibre alignment. The first broad type encompasses those particulates in which the longitudinal (chain) axis of individual cellulose nanocrystals is aligned in a single direction, whilst the second type exhibits more than one significant fibre orientation of nanocrystals within an aggregate. The TEM images of aerosol particulate types in **Figure 5** provide a useful legend. More specifically, aerosol particulate type 1 is a thin, high aspect-ratio fibrous particulate similar in morphology to never-sprayed cotton-derived CNCs typically observed in suspension. Particulate type 2 shows a similar morphology as type 1, albeit with greater width, and a more ellipsoidal shape, showing increased misalignment. Particulate type 3 is significantly larger and rounder, with a lower anisotropy and aspect ratio in two-dimensional space.

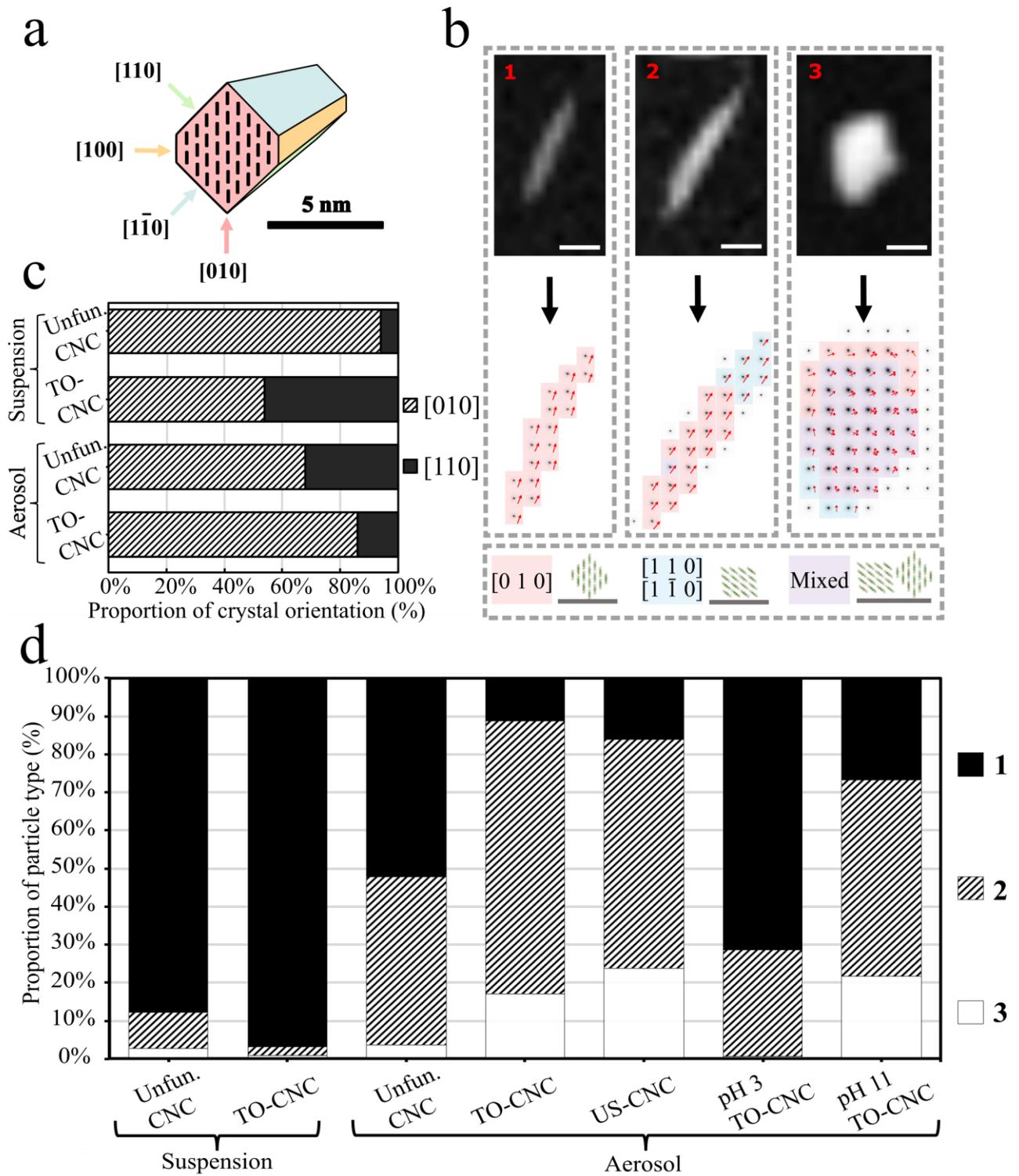

**Figure 5. (a)** Cross-sectional view of a cotton-isolated cellulose-Iβ crystal, labelled with its crystal zone-axes (normal to their corresponding crystal plane) **(b)** SNBED results showing the three different types of particulates observed in aerosolised CNCs and corresponding fibre axis orientation

maps **(c)** Relative proportion of crystal orientations in CNC suspensions and aerosols **(d)** Relative proportions of the particulate types in CNC suspensions and aerosols.

The SNBED results in **Figure 5b** show that both particle types 1 and 2 have a high degree of alignment of their component crystallites along their fibre (longitudinal) axis, which therefore coincides with the overall particulate axis. The interacting surface is predominately hydrophobic on the (1 0 0) plane in both aggregate types. A significant preferential lateral orientation is present, with the [0 1 0] axis (red pixels in **Figure 5b**) parallel to the electron beam. Considering the probe size is 15 nm, each diffraction pattern should contain structural information of at least a few crystallites along the lateral direction. The predominance of the [0 1 0] zone axis pattern thus indicates that the crystallites laterally aggregate through interactions between (0 1 0) surfaces of individual crystallites.

Similarly, type 2 aggregates show good uniaxial orientation, but occasionally with a lower degree of alignment compared to type 1. The lateral orientation is also similar to the case of type 1 particulates with (1 0 0) being the main interacting surface. However, as shown in **Figure 5b**, individual particulates often contain more than one lateral orientation, making the particulate structure "patchier" than that of particulate type 1.

SNBED analysis shows that central nanocrystals within the type 3 particulates possess a multilayered structure of CNCs possessing different orientations (**Figure 5b**). These electron diffraction (ED) patterns from the central area have higher inelastic scattering signal around the direct beam, indicating the comparatively large volume of material that makes up the aggregate. On the other hand, the outer CNCs of these aggregates often have a single orientation that traces only the contour lines of the aggregate morphology. These diffraction features suggest that the type 3 particulates have an approximately ellipsoidal cross-section. This is corroborated by atomic force microscopy (AFM) measurements (**Figure S14**). Furthermore, peripheral ED patterns imply that individual CNCs have assembled in a manner that minimises surface area. This is characteristic of droplet formation and subsequent evaporation. Thus, it is unlikely that particulate type 3 occurs as an artefact of sampling, whereby particulates would form on grids by piling up on individual CNCs in a random and less compact manner without a well-defined cross-sectional shape.

As shown in **Figure 5d**, in the instance of aerosolised TEMPO-oxidised CNCs (as well as aerosolised ultrasonicated CNCs and aerosolised TEMPO-oxidised CNCs at pH 11), particulate type 3 is present in significant number, at approximately 20% of the total particulate number. This is in comparison to less than 4% in aerosolised Unfun. CNCs and less than 3% in liquid Unfun. CNC and TO-CNC samples. Particulate type 2 also occurs in greater numbers in the aerosolised TO-CNC sample compared to the aerosolised Unfun. CNC sample.

Now, consider the case of Unfun. CNC aerosols, where surface charges and colloidal stability in liquid suspension are lower, significantly lower degrees of aggregation were observed upon aerosolisation. The overwhelming majority of particulates were type 1 and 2, with type 3 being almost negligible. The structures in particulate types 1 and 2 appeared almost identical to the corresponding types observed in the aerosolised TO-CNCs. However, higher frequencies of [1 1 0] and [1 -1 0] patterns were observed, indicating the differing interaction patterns between CNCs and TO-CNCs **(Figure 5c)**.

## 4. Conclusions

This work investigates the morphology of aerosolised CNCs using both inline analysis and offline imaging. For all the analysed samples, significant differences in the aspect ratio of aerosolised cellulose nanocrystals were observed in comparison to never-sprayed suspensions. Particle length appeared less susceptible to change during aerosolization, implying an emphasis on lateral aggregation as a driver of morphological change. The particle width and aspect ratio of aerosolised particle aggregates were highly influenced by surface charges: aerosol particle aggregation was shown to increase drastically with the increasing extent of carboxylation of the CNCs and, hence, with increased suspension colloidal stability. Finally, we found that aerosol particle aggregation increased with increasing sonication dose applied to the CNCs. Interestingly, the morphologies of the aerosolised cellulose nanocrystal aggregates in both the case of increased surface charge and that of increased sonication dose were very similar. This suggests that the colloidal stability (or metastability on the time scale of the aerosolisation experiment) drives the aggregation mechanism. To investigate this further we use scanning nano-beam electron diffraction, pendant droplet and kinetic arrest analysis, alongside an investigation of the Péclet numbers associated with the systems. It was shown that CNCs within a colloidally stable droplet were able to rearrange and self-organise, rather than becoming enriched at the droplet surface, thus maintaining surface tension during drying, resisting droplet breakup and forming large, round aggregates. However, it remains to be investigated, in this specific case, how the nebulisation process can be used to further tune aerosol morphology, in terms of the use of differing nozzle types (e.g. ultrasonic, rotary and electrospray nozzles) and operating conditions (e.g. pressure and suspension concentration), as well as to what extent the droplet size distribution of the initial spray might constrain the phenomena observed.

The extent of aggregation has a significant impact on the final particle properties, and therefore their application-specific performance. We believe that our findings might have implications for the understanding how CNCs aggregate in different aerosol applications, including spray drying, painting, cosmetics, drug delivery and agriculture. Further importance is placed upon the results due to their relevance to ecological and human toxicology, as many commercial CNC products undergo a spray drying step. With cellulose nanocrystal production expanding and the potential for societal exposure to be high, it is important to understand the risk to environmental and human health. Aerosol morphology and aggregation behaviour are critical to understanding the biological uptake mechanisms by humans and the immune response by the body. The increase in overall size and the decrease in aspect ratio, observed most prominently in TEMPO-oxidised CNCs after aerosolisation, might therefore be a positive one, from a toxicological point of view, as many studies that have investigated rods, tubes and fibres of nanocellulose agree that increasing aspect ratio, increasing length and decreasing diameter are primary drivers of both cellular uptake and toxicity (Endes et al., 2015; Fenoglio et al., 2012; Visani de Luna et al., 2023; Wang et al., 2019; Zheng & Yu, 2016).

**Abbreviations**

**AFM** Atomic Force Microscopy

**AR** Aspect Ratio

**CNC** Cellulose Nanocrystal

**CPC** Condensation Particle Counter

**DMA** Differential Mobility Analyser

**DLS** Dynamic Light Scattering

**ED** Electron Diffraction

**FTIR** Fourier Transform Infrared

**MWCO** Molecular Weight Cut Off

**SEM** Scanning Electron Microscopy

**SMPS** Scanning Mobility Particle Sizer

**SNBED** Scanning Nanobeam Electron Diffraction

**TEM** Transmission Electron Microscopy

**TEMPO** (2,2,6,6-Tetramethylpiperidin-1-yl)oxyl

**TO-CNC** TEMPO-Oxidised Cellulose Nanocrystal

**Unfun. CNC** Unfunctionalised Cellulose Nanocrystal

**US-CNC** Ultrasonicated Cellulose Nanocrystal

**Acknowledgements and funding sources**

This work was funded by: The Yusuf and Farida Hamied Foundation (D.W); The Engineering and Physical Sciences Research Council (EPSRC EP/W031019/1 to R.P., B.F.P., S.V.); Hiroshima University WPI-SKCM$^2$ (B.F.P.). The authors would also like to thank Dr. Heather Greer (University of Cambridge) for her help in acquiring the TEM images (EPSRC EP/P030467/1).

**Data Availability Statement**

Original data relating to this publication is available from the University of Cambridge data repository (*DOI will be added upon acceptance*).

**Declaration of Interest**

All authors declare that there is no conflict of interest.

**Author contributions: CRediT**

Conceptualisation (D.W, S.V, C.G); Investigation (D.W, C.H, J.W), Formal analysis (D.W, J.W, Y.O), Methodology (DW), Writing – original draft (DW), Writing – review and editing (R.P, B.F.P, C.G, S.V, Y.O), Resources (R.F), Supervision (C.G, S.V).